\documentclass{JHEP3}
\usepackage[reqno]{amsmath}
\usepackage{amsfonts,amssymb}
\usepackage{epsfig}
\usepackage{graphicx}

\newcommand{\half}{\frac{1}{2}}

\newcommand{\Tr}{\operatorname{Tr}}
\renewcommand{\Re}{\operatorname{Re}}
\renewcommand{\Im}{\operatorname{Im}}

\newcommand{\err}[2]{\raisebox{-0.4ex}
{$\stackrel{\scriptstyle +#1}{\scriptstyle -#2}$}}

\newcommand{\psibar}{\overline{\psi}}

\newcommand{\gm}{\gamma_{\mu}}
\newcommand{\smn}{\sigma_{\mu\nu}}
\newcommand{\snl}{\sigma_{\nu\lambda}}
\renewcommand{\l}{\lambda}

\newcommand{\pslash}{\not\!p}
\newcommand{\qslash}{\not\!q}
\newcommand{\kslash}{\not\!K}
\newcommand{\Kslash}{\not\!K}
\newcommand{\Qslash}{\not\!Q}
\newcommand{\Kt}{\widetilde{K}}

\newcommand{\KdQ}{K\!\cdot\!Q}

\newcommand{\Zquark}{Z_2}
\newcommand{\Zgluon}{Z_3}

\newcommand{\z}{{(0)}}

\newcommand{\csw}{c_{\text{sw}}}

\newcommand{\szz}{S_0^\z}
\newcommand{\siz}{S_I^\z}
\newcommand{\sizinv}[1]{S_I^\z(#1)^{-1}}

\newcommand{\gmomb}{g_{\overline{\text{MOM}}}}
\newcommand{\gmomt}{g_{\widetilde{\text{MOM}}}}

\newcommand{\MOMB}{$\overline{\text{MOM}}$}
\newcommand{\MOMT}{$\widetilde{\text{MOM}}$}

\newcommand{\momb}{\overline{\text{MOM}}}
\newcommand{\momt}{\widetilde{\text{MOM}}}
\newcommand{\msb}{\overline{\text{MS}}}

\newcommand{\cquarkII}{\cite{Skullerud:2001aw}}

\newcommand{\cqqg}{\cite{Skullerud:2002ge}}
\newcommand{\cqqgs}{\cite{Skullerud:PhD,Skullerud:1997wc,Skullerud:2002ge}}
\newcommand{\cdseggh}{\cite{vonSmekal:1998is,Atkinson:1998tu,Fischer:2003rp}}
\newcommand{\cdos}{\cite{Davydychev:2000rt}}

\title{Nonperturbative structure of the quark--gluon vertex}

\author{
Jonivar Skullerud\\
Instituut voor Theoretische Fysica, Universiteit van Amsterdam,
Valckenierstraat 65, NL--1018 XE Amsterdam, The Netherlands\\
E-mail: \email{jonivar@skullerud.name}
}
\author{Patrick O.\ Bowman, Ay{\c s}e K{\i}z{\i}lers{\"u}, Derek B.\
  Leinweber and Anthony G.\ Williams
\\
Special Research Centre for the Subatomic Structure of Matter, 
University of Adelaide, Adelaide SA 5005, Australia}

\abstract{The complete tensor structure of the quark--gluon vertex in
  Landau gauge is determined at two kinematical points (`asymmetric'
  and `symmetric') from lattice QCD in the quenched approximation.
  The simulations are carried out at $\beta=6.0$, using a mean-field
  improved Sheikholeslami--Wohlert fermion action, with two quark
  masses $\sim60$ and 115 MeV.  We find substantial deviations from
  the abelian form at the asymmetric point.  The mass dependence is
  found to be negligible.  At the symmetric point, the form factor
  related to the chromomagnetic moment is determined and found to
  contribute significantly to the infrared interaction strength.}

\preprint{ITFA 2003-13; ADP-03-108/T546}
\begin{document}

\section{Introduction}
\label{sec:intro}

The quark--gluon vertex describes the coupling between quarks and
gluons, and is thus one of the fundamental quantities of QCD.  In
perturbation theory, a complete calculation has been performed to one
loop \cdos, and partial two- and three-loop calculations have been
performed for specific gauges and kinematics
\cite{Chetyrkin:2000dq,Chetyrkin:2000fd}.  Nonperturbatively, however,
it remains largely unknown.  In \cqqgs\ the first steps were taken
towards a nonperturbative determination, by way of a quenched lattice
calculation of the form factor containing the running coupling in two
different kinematics in the Landau gauge.

The Dyson--Schwinger equation (DSE) for the quark propagator contains
the quark--gluon vertex, and normal practice has been to truncate the
hierarchy of DSEs by providing an {\em ansatz} for the vertex.
However, if a realistic gluon propagator, obtained from the coupled
ghost--gluon(--quark) DSEs \cdseggh\ and consistent with lattice data
\cite{Leinweber:1998uu,Bonnet:2000kw,Bonnet:2001uh} is used, dynamical
chiral symmetry breaking appears to be quite sensitive to the details
of the {\em ans{\"a}tze} employed \cite{Fischer:2003rp}.  It therefore
appears highly desirable to obtain `hard' information about the full
infrared structure, not only the part containing the running coupling.

In this paper we take the first steps towards this aim, by determining
all the nonzero form factors at the two kinematic points used in
\cqqg, namely $q=0$ and $q=-2p$, where $q$ is the gluon momentum and
$p$ is the momentum of the outgoing quark leg.  At the same time we
also study the quark mass dependence by using two different quark
masses for the vertex at $q=0$.  Some preliminary results have already
been presented in \cite{Skullerud:2002sk}.

The quark--gluon vertex is related to the ghost self-energy through
the Slavnov--Taylor identity,
\begin{equation}
q^\mu\Gamma_\mu(p,q) = G(q^2)
 \left[(1-B(q,p+q))S^{-1}(p) - S^{-1}(p+q)(1-B(q,p+q))\right] \, ,
\label{eq:sti}
\end{equation}
where $G(q^2)$ is the ghost renormalisation function and $B(q,k)$ is
the ghost--quark scattering kernel.  Evidence from lattice simulations
\cite{Bloch:2002we} and Dyson--Schwinger equation studies \cdseggh\
indicate that $G(p^2)$ is strongly infrared enhanced, and this should
also show up in the quark--gluon vertex.  On the other hand,
nontrivial structure in the ghost--quark scattering kernel, which has
usually been assumed to be small, may also be realised in the vertex.

The rest of the paper is structured as follows: In section
\ref{sec:notation} we briefly present our notation and procedure,
referring to \cqqg\ for the details.  In section \ref{sec:asym} we
present results for the vertex at the asymmetric point and compare to
the abelian (quark--photon) vertex, which is completely determined by
the Ward--Takahashi identity at this point.  In section \ref{sec:sym}
we present results for the vertex at the symmetric point, including
the `chromomagnetic' form factor $\tau_5$.  Finally, in section
\ref{sec:outlook} we summarise our results and discuss prospects for
further work.  Some tree-level formulae used in the analysis are given
in the Appendix.

\section{Notation and procedure}
\label{sec:notation}

Throughout this article, we will be using the same notation as in
\cqqg, and we refer to that article for a detailed discussion of our
notation and procedure.  We write the one-particle irreducible
(proper) vertex (see fig.~\ref{fig:vtx-ill}) as $\Lambda^a_\mu(p,q)
\equiv t^a\Lambda_\mu(p,q)$, where $p$ and $q$ are the outgoing quark
and gluon momentum respectively.  The incoming quark momentum is
denoted $k$.
\FIGURE{
\includegraphics[width=5.2cm]{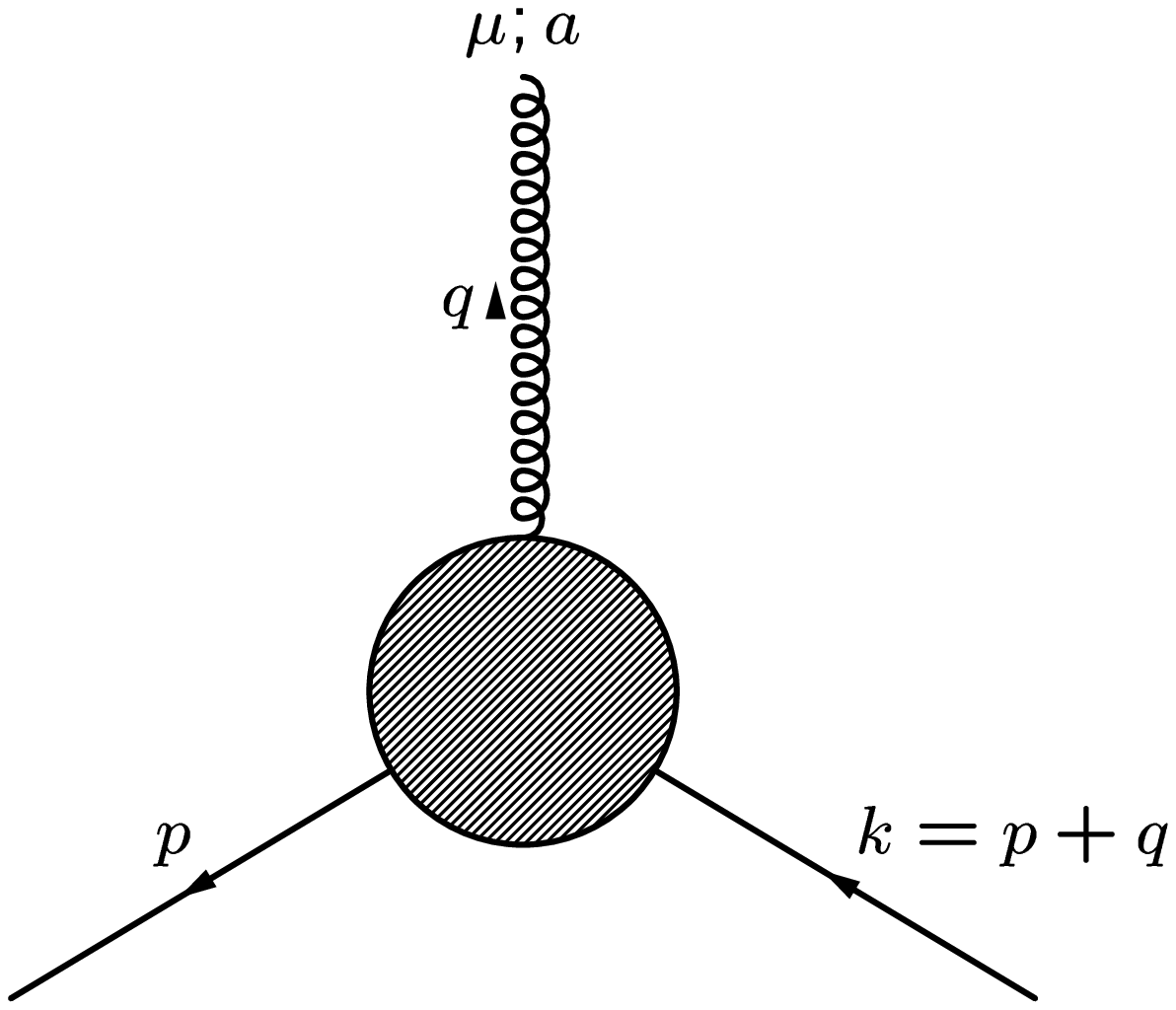}
\caption{The quark--gluon vertex.}
\label{fig:vtx-ill}
}

We will be operating in the Landau gauge, where, as discussed in
\cqqg, only the transverse-projected part of the vertex can be studied
away from $q=0$.  We will therefore define the transverse-projected
vertex as
\begin{equation}
 \Lambda^P_\mu(p,q) \equiv P_{\mu\nu}(q)\Lambda_\nu(p,q)
 = \Bigl(\delta_{\mu\nu} - \frac{q_\mu q_\nu}{q^2}\Bigr)
  \Lambda_\nu(p,q) \, .
\end{equation}
In a general kinematics the vertex can be decomposed
into 12 independent vectors which we can write in terms of vectors
$L_i,T_i$ and scalar functions $\lambda_i,\tau_i$ as described in
\cqqg: 
\begin{equation} 
\Lambda_\mu(p,q) = 
 -ig\sum_{i=1}^{4}\lambda_i(p^2,q^2,k^2)L_{i,\mu}(p,q) 
    -ig\sum_{i=1}^{8}\tau_i(p^2,q^2,k^2)T_{i,\mu}(p,q) \, .
\label{eq:decompose}
\end{equation}

We will here be focusing on the two specific kinematics defined in
\cqqg\ and related there to the \MOMT\ and \MOMB\ renormalisation
schemes --- namely, the `asymmetric' point $q=0$ (i.e., $p^2=k^2;
q^2=0$) and the `symmetric' point $q=-2p$ (i.e., $p^2=k^2=q^2/4$).  In
the asymmetric kinematics, the vertex reduces to
\begin{align}
\Lambda_\mu(p,0) =& -ig\Bigl[\lambda_1(p^2,0,p^2)\gm 
 - 4\lambda_2(p^2,0,p^2)\pslash p_\mu 
 - 2i\lambda_3(p^2,0,p^2)p_\mu\Bigr] \, , \label{eq:decomp-asym} \\
\intertext{while in the symmetric kinematics we have}
\begin{split}
\Lambda_\mu(-q/2,q) =& -ig\Bigl[\lambda_1(q^2/4,q^2,q^2/4)\gm
 + \tau_3(q^2/4,q^2,q^2/4)\bigl(\qslash q_\mu - q^2\gm\bigr) \\
&\phantom{-ig_0\Bigl[} - i\tau_5(q^2/4,q^2,q^2/4)\smn q_\nu \Bigr] \, ;
\end{split} \label{eq:decomp-sym} \\
\Lambda^P_\mu(-q/2,q) =& -ig\Bigl[
 \lambda'_1(q^2/4,q^2,q^2/4)\bigl(\gm-\qslash q_\mu/q^2\bigr) 
 - i\tau_5(q^2/4,q^2,q^2/4)\smn q_\nu\Bigr] \, ,
 \label{eq:decomp-sym-proj}
\end{align}
where on the last line, in the transverse-projected vertex, we have
written $\lambda'_1 \equiv \lambda_1-q^2\tau_3$.

In an abelian theory (QED), the Ward--Takahashi identities imply that
the form factors $\lambda_i (i=1,2,3)$ are given uniquely in terms of
the fermion propagator,
\begin{equation}
S(p) = \frac{1}{i\pslash A(p^2) + B(p^2)}\, = \, \frac{Z(p^2)}{i\pslash +
M(p^2)} \, .
\end{equation}
In the kinematics we are considering, they are given by
\begin{gather}
\lambda_1^{\text{QED}}(p^2,0,p^2)
 = \lambda_1^{\text{QED}}(p^2,4p^2,p^2) = A(p^2) \, ; \label{eq:bc1} \\
\lambda_2^{\text{QED}}(p^2,0,p^2) = 
 \half\frac{\mathrm{d}}{\mathrm{d}p^2}A(p^2) \, ; \qquad 
\lambda_3^{\text{QED}}(p^2,0,p^2) =
 -\frac{\mathrm{d}}{\mathrm{d}p^2}B(p^2) \, .\label{eq:bc23}
\end{gather}
The deviation of the QCD form factors from these expressions thus give
us a measure of the purely nonabelian nature of the theory.  Note that
$\lambda_4$, which is identically zero in QED, is zero also in QCD at
these two particular kinematic points.

The bare (unrenormalised) quantities $\lambda_3, \tau_5$ and
$\lambda'_1$ (at the symmetric point) can be obtained by tracing the
lattice $\Lambda_\mu$ with the appropriate Dirac matrix (the identity,
$\smn$ and $\gm$ respectively):
\begin{align}
\lambda_3(p^2,0,p^2) &= \frac{1}{2p^2}\sum_\mu
 p_\mu\frac{1}{4g_0}\Re\Tr\Lambda_\mu(p,0) \, ; \label{eq:l3} \\
\tau_5(q^2/4,q^2,q^2/4) &=
 -\frac{1}{3q^2}\sum_{\mu,\nu}
  q_\mu\frac{1}{4g_0}\Re\Tr\smn\Lambda^P_\nu(-q/2,q) \, ; \label{eq:t5} \\
\lambda'_1(q^2/4,q^2,q^2/4) &=
 -\frac{1}{3}\sum_\mu\frac{1}{4g_0}\Im\Tr\gm\Lambda^P_\mu(-q/2,q) \, .
\label{eq:l1sym}
\end{align}
At the asymmetric point, $\lambda_1$ and $\lambda_2$ both come with
the same Dirac structure.  To separate them, we first determine
$\lambda_1$ as described in \cqqg\ by setting the `longitudinal'
momentum component $p_\mu$ to zero, and then obtain $\lambda_2$ by
\begin{equation}
\lambda_2(p^2,0,p^2) = 
\frac{1}{4p^2}\sum_\mu\Bigl(\frac{1}{4g_0}\Im\Tr\gm\Lambda_\mu(p,0)
 + \lambda_1(p^2,0,p^2) \Bigr) \, .
\label{eq:l2}
\end{equation}

In order to make the lattice form factors more continuum-like, we
employ tree-level correction, as discussed in
\cite{Bonnet:2000kw,Skullerud:2001aw}.  The tree-level correction of
$\lambda_1$ is described in \cqqg, although at the symmetric point we
have here refined the correction procedure, as described in
appendix~\ref{sec:tree}.  In the case of $\lambda_2, \lambda_3$ and
$\tau_5$, these are simply zero at tree level in the continuum, while
they are non-zero on the lattice with the action and parameters we are
using.  We therefore have to subtract off the lattice tree-level
forms.  The details of this are given in appendix~\ref{sec:tree}.  Not
unexpectedly, this procedure leads to large cancellations which make
our results unreliable at large momenta.

As always, the quantities obtained from the lattice are bare
(unrenormalised) quantities.  
The relation between renormalised and bare quantities is given
by
\begin{equation}
\psi^0 = \Zquark^{1/2}\psi\,; \qquad \psibar^0 = \Zquark^{1/2}\psibar\,;
\qquad A_\mu^0 = \Zgluon^{1/2}A_\mu\,; \qquad g_0 = Z_g g\, ; \qquad
\xi_0 = Z_3\xi\, ,
\label{eq:renorm-fields}
\end{equation}
where $\Zquark,\Zgluon,Z_g$ are the quark, gluon and vertex (coupling)
renormalisation constants respectively.  The renormalised quark and
gluon propagator and quark--gluon vertex are related to their bare
counterparts according to
\begin{gather}
S^{\text{bare}}(p;a) = \Zquark(\mu;a)S(p;\mu) \, ;
\qquad
D^{\text{bare}}(q^2;a) = \Zgluon(\mu;a)D(q^2;\mu) \, ; \\
\Lambda_\mu^{\text{bare}}(p,q;a)
 = Z_{1F}^{-1}(\mu;a)\Lambda_\mu(p,q;\mu) \, .
\label{eq:vertex-renorm}
\end{gather}
Renormalisation may be carried out in a
momentum subtraction scheme.  For the quantities computed at the
asymmetric point, we will use the \MOMT\ scheme defined in \cqqg\,
requiring that $\lambda_1(\mu^2,0,\mu^2)=1$; while for the quantities
at the symmetric point we will use a modification of the \MOMB\
scheme, requiring $\lambda'_1(\mu^2/4,\mu^2,\mu^2/4)=1$.  In both
cases we choose $\mu=2$ GeV as our renormalisation scale.  We can then
easily match our results on to perturbation theory in the ultraviolet,
using the associated (\MOMT\ or \MOMB) running coupling.

We use the same ensemble and parameters as in \cqqg.  The Wilson gauge
action is used at $\beta=6.0$ on a $16^3\times48$ lattice.  The Sommer
scale provides an inverse lattice spacing of 2.12 GeV.  The mean-field
improved SW action is adopted with off-shell improvement in the
associated propagators.  Further details may be found in \cqqg.  In
order to study the quark mass dependence of the vertex, we have used
two values for the hopping parameter, $\kappa=0.137$ and 0.1381,
corresponding to a bare quark mass $m\approx115$ and 60 MeV
respectively.

\section{Asymmetric point}
\label{sec:asym}

First, we investigate the mass dependence of the $\lambda_1$ form
factor, which was already studied in \cqqg.  Since in this paper we
are primarily concerned with the deviation from the abelian
(Ball--Chiu) form, we show, in figure~\ref{fig:l1z}, the quantity
$Z(p^2)\lambda_1(p^2,0,p^2)$, which in an abelian theory would be
constant.  The clear infrared enhancement observed in \cqqg\ is
confirmed, and we also see that the mass dependence of this quantity
is negligible.  The slight difference in $\lambda_1$
between the two masses observed in \cite{Skullerud:2002sk} is in other
words entirely due to the mass-dependence of the quark renormalisation
function.
\FIGURE{
\includegraphics*[width=10cm]{lambda1_z.eps}
\caption{The unrenormalised form factor $\lambda_1(p^2,0,p^2)$
  multiplied by the quark renormalisation function $Z(p^2)$, as a
function of $p$.  In an abelian theory, this would be a
  $p$-independent constant.}
\label{fig:l1z}
}

In order to compare our results with the abelian forms
(\ref{eq:bc23}), we have fitted the tree-level
corrected quark propagator \cquarkII\ to the following functional
forms \cite{Bowman:2002kn},
\begin{align}
Z(p^2) \equiv 1/A(p^2) &= k\Biggl(1-\frac{c^2}{a^2p^2 + l^2}\Biggr) 
\label{eq:zfunc} \, ; \\
aM(p^2) \equiv aB(p^2)/A(p^2) &= 
 c_m\frac{l_m^{2(\alpha-1)}}{(a^2p^2)^\alpha + l_m^{2\alpha}} + m_f \, ,
\label{eq:mfunc}
\end{align}
where $k, c, l, c_m, l_m, \alpha$ and $m$ are fit parameters.
The best fit values are given in table~\ref{tab:fitparams}.  When
comparing with the renormalised vertex we use the values obtained from
the quark propagator renormalised at 2 GeV, which amounts to dividing
the unrenormalised values by $Z(4{\rm GeV}^2)$.
\TABLE{
\begin{tabular}{c|rrrrrrr}\hline
$m$ (GeV) & $k$ & $c$ & $l$ & $c_m$ & $l_m$ & $\alpha$ & $m_f$ \\ \hline
60 & 1.075 & 0.218 & 0.326 & 0.0261 & 0.400 & 1.232 & 0.0258 \\
115 & 1.045 & 0.208 & 0.316 & 0.0357 & 0.484 & 1.361 & 0.0670 \\ \hline   
\end{tabular}
\caption{Fit parameters for best fits of the quark propagator to the
functional forms (\protect\ref{eq:zfunc}) and
(\protect\ref{eq:mfunc}).  All fits have been performed to data
surviving a cylinder cut with radius 1 unit of spatial momentum, up to
a maximum momentum of $pa=1.2$ for the lighter quark mass and 1.4 for
the heavier mass.}
\label{tab:fitparams}
}
From these fits, we can then derive the abelian form factors
(\ref{eq:bc23}).

We will also compare our results with the one-loop Euclidean-space
expressions,
\begin{align}
\begin{split}
\lambda_2^{\msb}(p^2,0,p^2)&=\frac{g^2}{16 \pi^2}\,\frac{1}{4p^2}\Bigg\{
 \left(1-2\frac{m^2}{p^2}\right)\left[2\xi\,C_F+\frac{C_A}{2}(1-\xi)\right] \\
& \phantom{=-\frac{g^2}{16 \pi^2}\,\frac{1}{4p^2}\Bigg\{}
  +\frac{m^4}{p^4}\ln\left(1+\frac{p^2}{m^2}\right)\,
    \left[4\xi\,C_F+(1-\xi)C_A\right]\Bigg\}
\end{split} \label{eq:l2-1loop} \, ;\\
\lambda_3^{\msb}(p^2,0,p^2)&=\frac{g^2}{16 \pi^2}\,\frac{m}{p^2}\Bigg\{
  \left[(3+\xi)C_F-(3+2\xi)\frac{C_A}{4}\right]
  \left(1-\frac{m^2}{p^2}\ln\left(1+\frac{p^2}{m^2}\right)\right)\Bigg\}
  \, ,
\label{eq:l3-1loop}
\end{align}
where the group factors $C_F=\frac{4}{3}$ and $C_A=3$ in QCD, and the
gauge parameter $\xi=0$ in Landau gauge.  In order to match this to
our lattice results, we renormalise both the lattice and perturbative
data in the \MOMT\ scheme.  From the data of fig.~6 in \cqqg\ we find
that $1/Z_{1F}^{\momt}(2{\rm GeV},a)=1.39\err{6}{7}$ at $m=115$ MeV.
From this we determine the renormalised form factors
$\lambda_{2,3}^{\momt} = Z_{1F}^{\momt}\lambda_{2,3}^{\rm lat}$.  The
\MOMT\ 1-loop values are determined by evaluating the expressions
(\ref{eq:l2-1loop}), (\ref{eq:l3-1loop}) using $\gmomt(2{\rm
GeV})=2.21(10)$ and multiplying by
$Z_{1F}^{\momt}/Z_{1F}^{\msb}=1.069$, obtained from eq.~(7.2) of
\cqqg.

In figure~\ref{fig:l2} we show the form factor $\lambda_2$ as a
function of $p$, for the heavier quark mass.  We see that it is
greatly enhanced both compared to the Ball--Chiu form (\ref{eq:bc23})
and the one-loop form (\ref{eq:l2-1loop}), and only approaches these
around or above 3 GeV.\footnote{In \cite{Skullerud:2002sk} there was
an error of a factor of 4 in the normalisation of $\lambda_2$, which
gave the false impression that our numerical results agree almost
perfectly with the Ball--Chiu form.}  
In figure~\ref{fig:l2p2} we show
the dimensionless quantity $4p^2\lambda_2(p^2,0,p^2)$ as a function of
$p$.  
\FIGURE{
\includegraphics*[width=10cm]{lambda2_ren_cutoff.eps}
\caption{The renormalised form factor $\lambda_2(p^2,0,p^2)$ as a
function of $p$.  Also shown is the abelian (Ball--Chiu) form of
(\protect\ref{eq:bc23}) and the one-loop form of
(\protect\ref{eq:l2-1loop}).}
\label{fig:l2}
}
\FIGURE{
\includegraphics*[width=10cm]{lambda2_p2_ren.eps}
\caption{The renormalised form factor $4p^2\lambda_2(p^2,0,p^2)$ as a
function of $p$.}
\label{fig:l2p2}
}
This quantity measures the relative strength of this component
compared to the tree-level $\lambda_1$.  We see that $\lambda_2$
becomes comparable in strength to $\lambda_1$ for the most infrared
points.

In figure~\ref{fig:l3} we show $\lambda_3(p^2,0,p^2)$ as a function of
$p$.  Here we have performed a `cylinder cut' \cite{Leinweber:1998im}
with radius 1 unit of spatial momentum to select data close to the
4-dimensional diagonal.  We see that it coincides within errors with
the Ball--Chiu form (\ref{eq:bc23}), and approaches the one-loop form
at about 2 GeV.  We also see that the quark mass dependence of both
$\lambda_2$ and $\lambda_3$ is very weak.
\FIGURE{
\includegraphics*[width=10cm]{lambda3_ren_cut.eps}
\caption{The renormalised form factor $\lambda_3(p^2,0,p^2)$ as a
function of $p$.  Also shown is the abelian (Ball--Chiu) form of
(\protect\ref{eq:bc23}) and the one-loop form of
(\protect\ref{eq:l3-1loop}).}
\label{fig:l3}
}
\FIGURE[b]{
\includegraphics*[width=10cm]{lambda3_p_ren.eps}
\caption{The renormalised form factor $\lambda_3(p^2,0,p^2)$
multiplied by twice the quark momentum $2p$, as a function of $p$, for
$m=115$ MeV.  This dimensionless quantity gives a measure of the
relative strength of $\lambda_3$.}
\label{fig:l3p}
}
$\lambda_3$ becomes somewhat larger as the
quark mass is decreased, which corresponds to the effect of dynamical
chiral symmetry breaking being relatively larger for a smaller bare
mass. 

In figure~\ref{fig:l3p} we show $2p\lambda_3(p^2,0,p^2)$ as a function
of $p$.  This quantity is dimensionless and measures the relative
strength of $\lambda_3$ compared to the tree-level $\lambda_1$.  For
the most infrared points, $\lambda_3$ can also be seen to contribute
significantly to the interaction strength, although clearly not as
strongly as $\lambda_1$ and $\lambda_2$.

In order to see more clearly the relative strength of all three
components of the vertex, in figure~\ref{fig:asym-all} we show the
dimensionless quantities $\lambda_1, 4p^2\lambda_2$ and $2p\lambda_3$
for the heavier quark mass.  In this figure, the hierarchy of
strengths $\l_1>\l_2>\l_3$ is evident.
\FIGURE{
\includegraphics*[width=10cm]{asym_all.eps}
\caption{The dimensionless form factors $\lambda_1, 4p^2\lambda_2$ and
$2p\lambda_3$ at the asymmetric point, as a function of $p$, for
$m=115$ MeV.}
\label{fig:asym-all}
}

\section{Symmetric point}
\label{sec:sym}

Since we have already established that the dependence of the vertex on
the quark mass is very weak, in this section we will only be using one
quark mass, $m\approx115$ MeV.  We will also in this section make use
of the lattice momentum variables $K(p)\equiv\sin(pa)/a$ and
$Q(q)\equiv2\sin(qa/2)/a=-2K(p)$.  These momentum variables appear in
the lattice tree-level expressions for the form factors we will be
studying, as well as in the transverse projector, and are thus
appropriate variables to use.

In figure~\ref{fig:l1sym}, we show $\lambda'_1$ at the symmetric point
as a function of $|Q(q)|$.  In contrast to in \cqqg, the tree-level
correction here has been carried out on each Lorentz component of the
vertex separately, as explained in the Appendix.  These results should
therefore be more reliable than those shown in \cqqg.  We have also
performed a cylinder cut on the data with a radius of 2 units of
spatial momentum in $q$. From these data, we determine
$1/Z_{1F}^{\momb}(2{\rm GeV},a) = 0.95(8)$.  Multiplying by $Z_2
Z_3^{1/2}$ determined in \cqqg\ we also find $\gmomb(2{\rm GeV}) =
1.47(15)$.  
\FIGURE{
\includegraphics*[width=10cm]{lambda1_sym_cut.eps}
\caption{The unrenormalised form factor $\lambda'_1(p^2,q^2,p^2)$ at
the symmetric point $q=-2p$, as a function of the gluon momentum $q$.
The data shown are those surviving a cylinder cut with radius 2 units
of spatial momentum in $q$.}
\label{fig:l1sym}
}
\FIGURE{
\includegraphics*[width=10cm]{lambda1_sym_ren.eps}
\caption{The renormalised form factor $\lambda'_1(p^2,q^2,p^2)$ at
the symmetric point $q=-2p$, as a function of the gluon momentum $q$.
Also shown is the one-loop form from \protect\cqqg.}
\label{fig:l1symren}
}
The ratio of renormalisation constants is
$Z_{1F}^{\momb}/Z_{1F}^{\msb} = 1.093$.  This is used to determine the
one-loop $\lambda'_1$, shown together with the renormalised lattice
$\lambda'_1$ in figure~\ref{fig:l1symren}.

In figure~\ref{fig:tau5-cut} we show the form factor $\tau_5$ as a
function of the gluon momentum $q$.  The same cylinder cut has been
performed as in fig.~\ref{fig:l1sym}.  We see that, although $\tau_5$
is power suppressed in the ultraviolet, it rises very significantly
for $q\lesssim2$ GeV.
\FIGURE{
\includegraphics*[width=10cm]{tau5_ren.eps}
\caption{The renormalised form factor $\tau_5$ at the symmetric point
as a function of the gluon momentum $q$.  The data shown are those
surviving a cylinder cut with radius 2 units of spatial momentum in
$q$.  Also shown is the one-loop form of
(\protect\ref{eq:tau5-1loop}).}
\label{fig:tau5-cut}
}
Although this form factor is
related to the chromomagnetic moment, and as such is expected to be of
phenomenological importance, it has not previously been included in
QED-inspired model vertices commonly used in, e.g., DSE-based studies.
However, work is in progress to provide an analytical, nonperturbative
expression for this and the other form factors in the purely
transverse part of the vertex \cite{Kizilersu:2003xx}.
We will also compare our lattice results to the one-loop $\tau_5$,
which in Euclidean space is given by
\begin{equation}
\begin{split}
\tau_5^{\msb}(&s^2,4s^2,s^2) = 
  \frac{g^2}{16\pi^2}\frac{m}{12s^2}\Biggl\{
 (1-\xi)\Bigl[8C_F + \xi C_A -
  C_A\frac{m^2(1-\xi)}{s^2-m^2}\Bigr] \\
& - (2C_F-C_A)\frac{4s^2(1-\xi)}{s^2+m^2}
  \biggl[\half\sqrt{1+\frac{m^2}{s^2}}
  \ln\frac{\sqrt{1+\frac{m^2}{s^2}}+1}{\sqrt{1+\frac{m^2}{s^2}}-1}
  + \half\ln\frac{m^2}{\mu^2}\biggr] \\
&+ C_A\frac{s^2}{s^2-m^2}
 \biggl[7+9\xi+2\xi^2-\frac{2m^2(1-\xi)}{s^2-m^2}
  \Bigl(6-\xi+\frac{2m^2(1-\xi)}{s^2-m^2}\Bigr)\biggr]
  \ln\frac{4s^2}{\mu^2} \\
& + \biggl[4C_F(1-\xi)\frac{s^2-2m^2}{s^2+m^2}
 - C_A\Bigl(9+7\xi+2\xi^2 - \frac{6m^2(1-\xi)}{s^2+m^2} \\
&\phantom{+\biggl[4C_F} - (6-16\xi+\xi^2)\frac{2m^2}{s^2-m^2}
 - \frac{2m^4(1-\xi)}{(s^2-m^2)^2}
\bigl[9-4\xi+\frac{2m^2(1-\xi)}{s^2-m^2}\bigr]\Bigr)
\biggr] \times \\
& \qquad\times\biggl[\ln\frac{s^2+m^2}{\mu^2}
  +\frac{m^2}{s^2}\ln\Bigl(1+\frac{s^2}{m^2}\Bigr)\biggr]
\Biggr\} \, .
\end{split}
\label{eq:tau5-1loop}
\end{equation}
We find that the nonperturbative $\tau_5$ is several orders of
magnitude larger than the one-loop form, and there is no sign of the
lattice data approaching the perturbative form even for the most
ultraviolet points we can trust, around 5 GeV.  We take this as an
indication that very strong nonperturbative effects affect this form
factor.  It is also worth noting that the one-loop contribution to
both $\tau_5$ and $\lambda'_1$ at the symmetric point are an order of
magnitude smaller than the one-loop contributions to form factors at
the asymmetric point.

In order to get a dimensionless measure of the strength of this
component relative to the tree-level vertex, we have scaled $\tau_5$
with the gluon momentum $q$.  We show this together with $\lambda'_1$
in figure~\ref{fig:sym-all}.  As we can see, between 1 and 2 GeV,
$\tau_5$ contributes with about the same strength as $\lambda'_1$,
making it a very significant contribution that cannot be ignored.
\FIGURE[t]{
\includegraphics*[width=8.8cm]{sym_all.eps}
\caption{The dimensionless form factors $\lambda'_1$ and $q\tau_5$ at
the symmetric point, as a function of $q$.  These quantities gives a
measure of the relative strength of the two components of the vertex.}
\label{fig:sym-all}
}

Although $\tau_5$ has the same tensor structure as the
(chromo-)magnetic moment, the relation between the two is not
straightforward.  In particular, since quarks are never on-shell, the
Gordon decomposition which is used to define the magnetic moment in
QED is not applicable, making the definition of the chromomagnetic
moment ambiguous.  This is an issue that deserves further
investigation.

\section{Outlook}
\label{sec:outlook}

We have computed the complete quark--gluon vertex at two kinematical
points, finding substantial deviations from the abelian form --- which
cannot be described by a universal function multiplying the abelian
form as in \cite{Fischer:2003rp}.  This, and the fact that we observe
a $p$-dependent enhancement of $\lambda_1$ at the asymmetric point,
where $q$ and thereby also the ghost form factor $G(q^2)$ is fixed,
indicates that the ghost--quark scattering kernel entering into the
Slavnov--Taylor identity (\ref{eq:sti}) must contain nontrivial
structure.

The form factor $\tau_5$, related to the chromomagnetic moment, has
been estimated nonperturbatively for the first time, and found to be
important.  The work has been carried out on a relatively small
lattice, using a fermion discretisation which has serious
discretisation errors at large momenta.  It will be important to
repeat this study using larger lattices and a more well-behaved
fermion discretisation.

A natural extension of this work would be to map out the entire
kinematical space in the three variables $p^2, k^2, q^2$.  This is
numerically very demanding, but work is underway on a complete
determination of $\lambda'_1(p^2,q^2,k^2)$.

Finally, it should be noted that the lattice Landau gauge restricts us
to computing only the transverse-projected vertex away from $q=0$;
i.e., it is not possible to determine $\lambda_i, \tau_i,
i=1,\ldots,4$ separately; only the linear combinations $\lambda'_i$.
Although the vertex is always contracted with the gluon propagator in
all actual applications, and thus only the transverse-projected vertex
plays any role in Landau gauge, it would be of interest to determine
all these form factors by computing the vertex in a general covariant
gauge --- which would also give a handle on the important issue of
gauge dependence.

\section*{Acknowledgments} 

This work has been supported by Stichting FOM and the Australian
Research Council.  We acknowledge the use of UKQCD configurations for
this work.

\appendix
\section{Tree-level expressions}
\label{sec:tree}

The tree-level lattice expressions are given in terms of the lattice
momentum variables,
\begin{align}
K_\mu(p) & \equiv  \frac{1}{a}\sin(p_\mu a) \label{def:lat-K} \,; \\
Q_\mu(p) &
 \equiv  \frac{2}{a}\sin(p_{\mu}a/2) = \frac{\sqrt{2}}{a}\sqrt{1-\cos(p_\mu a)} 
\label{def:lat-Q} \,; \\
\Kt_\mu(p) & \equiv \half K_\mu(2p) = \frac{1}{2a}\sin(2p_\mu a)
\label{def:lat-Kt} \,; \\
C_\mu(p) & \equiv \cos(p_\mu a) \,.\label{def:lat-C}
\end{align}
The tree-level vertex is \cqqg
\begin{equation}
\Lambda^{a(0)}_{I,\mu}(p,q) =
c_m\sizinv{p}\szz(p)\Lambda^{a(0)}_{0,\mu}(p,q) 
\szz(p+q)\sizinv{p+q} \, ,
\label{eq:tree-imp}
\end{equation}
and
\begin{align}
(\siz)^{-1}\szz & = \frac{1}{D_I}\Bigl(i\kslash A_V + B_V\Bigr) \, ,
\label{def:ab}
\end{align}
where we have written
\begin{align}
c_m &\equiv 1+b_qm \, ; \\
A_V(p) &= 2c'_qD(p) \, ; \\
B_V(p) &= (c_m-2c'_qM(p))D(p) \, ; \\
D_I(p) &= \bigl[A_V^2(p)K^2(p) + B_V^2(p)\bigr]/D(p) \, ; \\
D(p) &= K^2(p) + M^2(p) \, ; \\
M(p) &= m + \half Q^2(p) \, .
\end{align}

At $q=0$ we have
\begin{equation}
\Lambda^{a(0)}_{I,\mu}(p,0) =
\frac{-ig_0}{D_I^2}\Bigl(i\kslash A_V(p) + B_V(p)\Bigr)
\Bigl(\gm C_\mu(p) - iK_\mu(p)\Bigr)
\Bigl(i\kslash A_V(p) + B_V(p)\Bigr) \, .
\end{equation}
This expands to
\begin{equation}
\begin{split}
\Bigl(i\kslash&A_V + B_V\Bigr)\Bigl(\gm C_\mu - iK_\mu\Bigr)
\Bigl(i\kslash A_V + B_V\Bigr) \\
= & (A_V^2K^2 + B_V^2)\gm C_\mu - 2A_V^2C_\mu \kslash K_\mu
  + 2iA_VB_VK_\mu C_\mu\\ & - iB_V^2K_\mu + 2A_VB_V\kslash K_\mu
  + iA_V^2K^2K_\mu \\
= & (A_V^2K^2 + B_V^2)C_\mu\gm + 2(A_VB_V-A_V^2C_\mu)\kslash K_\mu \\
 & + i(2A_VB_VC_\mu + A_V^2K^2 - B_V^2)K_\mu \, .
\end{split}
\end{equation}

The tree-level form factors $\lambda_2^\z, \lambda_3^\z$ can be read
off directly:
\begin{align}
\l_2^\z + \tilde{\l}_2^\z C_\mu
 &= \frac{c_m}{2D_I^2}\Bigl(A_VB_V-A_V^2C_\mu\Bigr)
  = c_mc'_q\frac{D^2}{D_I^2}\Bigl[c_m-2c'_qM-2c'_qC_\mu\Bigr] \, ;\\
\l_3^\z + \tilde{\l}_3^\z C_\mu
 &= \frac{c_m}{2D_I^2}\Bigl(A_V^2K^2-B_V^2 + 2A_VB_VC_\mu\Bigr) \, .
\end{align}
The lattice, tree-level corrected equivalents of (\ref{eq:l2}) and
(\ref{eq:l3}), which we use to obtain $\lambda_2$ and $\lambda_3$, are
thus 
\begin{align}
\begin{split}
\lambda_2(p^2,0,p^2) &= 
\frac{1}{4K^2(p)}\sum_\mu\biggl[\frac{1}{4g_0}\Im\Tr\gm\Lambda_\mu(p,0)
 + \lambda_1(p^2,0,p^2) \\
& \phantom{=\frac{1}{4K^2(p)}\sum_\mu\biggl[}
 - 4K_\mu^2(p)\Bigl(\l_2^\z(p) +\tilde{\l}_2^\z(p)C_\mu(p)\Bigl)\biggr]
 \, ; \end{split}\label{eq:l2lat} \\
\lambda_3(p^2,0,p^2) &= \frac{1}{2K^2(p)}\sum_\mu \biggl[
 K_\mu(p)\frac{1}{4g_0}\Re\Tr\Lambda_\mu(p,0)
 - 2K_\mu^2(p)\Bigl(\l_3^\z(p) + \tilde{\l}_3^\z(p)C_\mu(p)\Bigr)\biggr]
  \, . \label{eq:l3lat}
\end{align}

The tree-level vertex at the symmetric point is given by eq.~(B.21) of
\cqqg.  We use the following decomposition into independently
transverse tensors,
\begin{equation}
\begin{split}
\frac{i}{g_0}\Lambda^\z_{I,\mu}(p,-2p) =& \lambda_1^\z\gamma_\mu
 - \tau_3^\z(Q^2\gamma_\mu-\Qslash Q_\mu)
 - \tilde{\tau}_3^\z(\KdQ C_\mu\gamma_\mu-\Kslash K_\mu) \\
& -i\tau_5^\z\sum_\nu\smn Q_\nu
 -i\tilde{\tau}_5^\z C_\mu\sum_\nu\smn K_\nu \\
& -i\tilde{\tau'}_5^\z\Bigl[C_\mu\sum_\nu\smn Q_\nu
   + K_\mu\sum_{\nu\lambda}\snl Q_\nu K_\lambda/(Q\cdot K)\Bigr] \, ,
\end{split}
\end{equation}
where $Q\equiv Q(q), K\equiv K(q), C\equiv C(q/2)$; $\l_1^\z,
\tau_3^\z$ and $\tilde{\tau}_3^\z$ are given by (B.25)--(B.27) of
\cqqg, and
\begin{align}
\tau_5^\z &= \frac{c_m}{D_I^2}{A_V(p)B_V(p)}
 = 2c_mc'_q(c_m-2c'_q M(p))\frac{D^2(p)}{D_I^2(p)} \, ; \\
\tilde{\tau}_5^\z & = -\frac{\csw c_m}{2}\frac{D(p)}{D_I(p)} \, ; \\
\tilde{\tau'}_5^\z &
 = \csw c_m \frac{A_V^2(p)(K(p)\cdot\Kt(p))}{D_I^2(p)}
 = 4\csw c_m {c'_q}^2(K(p)\cdot\Kt(p))\frac{D^2(p)}{D_I^2(p)} \, .
\end{align}
Since the continuum $\lambda'_1(\gm-\qslash q_\mu/q^2)$ becomes two
independent tensors on the lattice,
\begin{equation}
\lambda'_1\left(\frac{q^2}{4},q^2,\frac{q^2}{4}\right)
\left(\gm - \frac{\qslash q_\mu}{q^2}\right) \to
 (\lambda_1 - Q^2\tau_3)\left(\gm - \frac{\Qslash Q_\mu}{Q^2}\right)
 - \tilde{\tau}_3(\KdQ C_\mu\gamma_\mu-\Kslash K_\mu) \, ,
\end{equation}
we cannot simply factor out the tree-level behaviour with a simple
multiplicative correction.  Instead we apply a `hybrid' scheme where
the dominant term, multiplying $(\gm - \Qslash Q_\mu/Q^2)$, is
corrected multiplicatively, after first subtracting off the remaining
part,
\begin{equation}
\tilde{\tau}_3^\z\Bigl[(\KdQ C_\mu\gamma_\mu-\Kslash K_\mu) 
 - \KdQ C_\mu(\gm - \Qslash Q_\mu/Q^2)\Bigr]
 = -\tilde{\tau}_3^\z\Bigl(K_\mu^2-K_\mu Q_\mu\frac{\KdQ}{Q^2}\Bigr)
\, .
\end{equation}
It turns out that this term is completely negligible, but it has still
been included in the correction.  Thus, the lattice, tree-level
corrected equivalent of (\ref{eq:l1sym}) which we use to compute
$\l'_1$, is
\begin{align}
\lambda'_1\left(\frac{q^2}{4},q^2,\frac{q^2}{4}\right) =&
 \frac{1}{3}\sum_\mu\frac{-\frac{1}{4g_0}\Im\Tr\gm\Lambda^P_\mu(-q/2,q)
  -\tilde{\tau}_3^\z\Bigl(K_\mu^2-K_\mu Q_\mu\frac{\KdQ}{Q^2}\Bigr)}
 {\l_1^\z - Q^2\tau_3^\z - \KdQ\tilde{\tau}_3^\z C_\mu} \, .
  \label{eq:l1sym-lat} 
\end{align}
For $\tau_5$ we employ an additive correction scheme, and thus the
lattice equivalent of (\ref{eq:t5}) is
\begin{align}
\begin{split}
\tau_5\left(\frac{q^2}{4},q^2,\frac{q^2}{4}\right) =&
 -\frac{1}{3Q^2(q)}\sum_{\mu,\nu}\biggl[
  Q_\mu(q)\frac{1}{4g_0}\Re\Tr\smn\Lambda^P_\nu(-q/2,q)  \\
& - Q_\mu^2\Bigl\{\tau_5^\z + C_\mu C_\nu\tilde{\tau}_5^\z
  + C_\nu\tilde{\tau'}_5^\z
   \bigl[1-(C_\nu-C_\mu)Q_\nu^2/Q\cdot\!\!K\bigr]\Bigr\}\biggr]
\, .
\end{split}
  \label{eq:t5lat}
\end{align}


\begin{thebibliography}{10}

\bibitem{Davydychev:2000rt}
A.~I. Davydychev, P.~Osland and L.~Saks, {\it Quark gluon vertex in arbitrary
  gauge and dimension},  {\em Phys. Rev.} {\bf D63} (2001) 014022
  [\href{http://arXiv.org/abs/hep-ph/0008171}{{\tt hep-ph/0008171}}].

\bibitem{Chetyrkin:2000dq}
K.~G. Chetyrkin and A.~R{\'e}tey, {\it Three-loop three-linear vertices and
  four-loop $\widetilde{MOM}$ $\beta$ functions in massless {QCD}},
  \href{http://arXiv.org/abs/hep-ph/0007088}{{\tt hep-ph/0007088}}.

\bibitem{Chetyrkin:2000fd}
K.~G. Chetyrkin and T.~Seidensticker, {\it Two loop {QCD} vertices and three
  loop {MOM} $\beta$ functions},  {\em Phys. Lett.} {\bf B495} (2000) 74--80
  [\href{http://arXiv.org/abs/hep-ph/0008094}{{\tt hep-ph/0008094}}].

\bibitem{Skullerud:PhD}
J.~I. Skullerud, {\em Renormalisation in lattice {QCD}}.
\newblock PhD thesis, University of Edinburgh, 1996.

\bibitem{Skullerud:1997wc}
{\bf UKQCD} Collaboration, J.~I. Skullerud, {\it The running coupling from the
  quark gluon vertex},  {\em Nucl. Phys. Proc. Suppl.} {\bf 63} (1998) 242
  [\href{http://arXiv.org/abs/hep-lat/9710044}{{\tt hep-lat/9710044}}].

\bibitem{Skullerud:2002ge}
J.~Skullerud and A.~K{\i}z{\i}lers{\"u}, {\it Quark-gluon vertex from lattice
  {QCD}},  {\em JHEP} {\bf 09} (2002) 013
  [\href{http://arXiv.org/abs/hep-ph/0205318}{{\tt hep-ph/0205318}}].

\bibitem{vonSmekal:1998is}
L.~von Smekal, A.~Hauck and R.~Alkofer, {\it A solution to coupled
  {D}yson-{S}chwinger equations for gluons and ghosts in {L}andau gauge},  {\em
  Ann. Phys.} {\bf 267} (1998) 1
  [\href{http://arXiv.org/abs/hep-ph/9707327}{{\tt hep-ph/9707327}}].

\bibitem{Atkinson:1998tu}
D.~Atkinson and J.~C.~R. Bloch, {\it Running coupling in non-perturbative
  {QCD}. {I}: Bare vertices and y-max approximation},  {\em Phys. Rev.} {\bf
  D58} (1998) 094036 [\href{http://arXiv.org/abs/hep-ph/9712459}{{\tt
  hep-ph/9712459}}].

\bibitem{Fischer:2003rp}
C.~S. Fischer and R.~Alkofer, {\it Non-perturbative propagators, running
  coupling and dynamical quark mass of {L}andau gauge {QCD}},
  \href{http://arXiv.org/abs/hep-ph/0301094}{{\tt hep-ph/0301094}}.

\bibitem{Leinweber:1998uu}
{\bf UKQCD} Collaboration, D.~B. Leinweber, J.~I. Skullerud, A.~G. Williams and
  C.~Parrinello, {\it Asymptotic scaling and infrared behavior of the gluon
  propagator},  {\em Phys. Rev.} {\bf D60} (1999) 094507
  [\href{http://arXiv.org/abs/hep-lat/9811027}{{\tt hep-lat/9811027}}].

\bibitem{Bonnet:2000kw}
F.~D.~R. Bonnet, P.~O. Bowman, D.~B. Leinweber and A.~G. Williams, {\it
  Infrared behavior of the gluon propagator on a large volume lattice},  {\em
  Phys. Rev.} {\bf D62} (2000) 051501
  [\href{http://arXiv.org/abs/hep-lat/0002020}{{\tt hep-lat/0002020}}].

\bibitem{Bonnet:2001uh}
F.~D.~R. Bonnet, P.~O. Bowman, D.~B. Leinweber, A.~G. Williams and J.~M.
  Zanotti, {\it Infinite volume and continuum limits of the {L}andau-gauge
  gluon propagator},  {\em Phys. Rev.} {\bf D64} (2001) 034501
  [\href{http://arXiv.org/abs/hep-lat/0101013}{{\tt hep-lat/0101013}}].

\bibitem{Skullerud:2002sk}
J.~Skullerud, P.~Bowman and A.~K{\i}z{\i}lers{\"u}, {\it The nonperturbative
  quark gluon vertex},  \href{http://arXiv.org/abs/hep-lat/0212011}{{\tt
  hep-lat/0212011}}.

\bibitem{Bloch:2002we}
J.~C.~R. Bloch, A.~Cucchieri, K.~Langfeld and T.~Mendes, {\it Running coupling
  constant and propagators in {SU(2)} {L}andau gauge},
  \href{http://arXiv.org/abs/hep-lat/0209040}{{\tt hep-lat/0209040}}.

\bibitem{Skullerud:2001aw}
J.~Skullerud, D.~B. Leinweber and A.~G. Williams, {\it Nonperturbative
  improvement and tree-level correction of the quark propagator},  {\em Phys.
  Rev.} {\bf D64} (2001) 074508
  [\href{http://arXiv.org/abs/hep-lat/0102013}{{\tt hep-lat/0102013}}].

\bibitem{Bowman:2002kn}
P.~O. Bowman, U.~M. Heller, D.~B. Leinweber and A.~G. Williams, {\it Modelling
  the quark propagator},  \href{http://arXiv.org/abs/hep-lat/0209129}{{\tt
  hep-lat/0209129}}.

\bibitem{Leinweber:1998im}
{\bf UKQCD} Collaboration, D.~B. Leinweber, J.~I. Skullerud, A.~G. Williams and
  C.~Parrinello, {\it Gluon propagator in the infrared region},  {\em Phys.
  Rev.} {\bf D58} (1998) 031501
  [\href{http://arXiv.org/abs/hep-lat/9803015}{{\tt hep-lat/9803015}}].

\bibitem{Kizilersu:2003xx}
A.~K{\i}z{\i}lers{\"u} and M.~R. Pennington, ``Building the full fermion-boson
  vertex of {QED} by imposing the multiplicative renormalizability of the
  {S}chwinger-{D}yson equations for the fermion and boson propagators.'' In
  preparation, 2003.

\end{thebibliography}
\providecommand{\href}[2]{#2}\begingroup\raggedright\endgroup

\end{document}